%% file: ms.tex
\documentclass[12pt]{amsart}
\newcommand{\mum}{\mbox{$\mu$m}}

\newcommand{\hcop}{\mbox{HCO$^+$}}

\newcommand{\kms}{\mbox{km s$^{-1}$}}

\newcommand{\micron}{{$\mu$m}}

\newcommand\arcdeg{\mbox{$^\circ$}}%
\newcommand\arcsec{\mbox{$^{\prime\prime}$}}%
\newcommand\ion[2]{#1$\;${\small\rmfamily\@Roman{#2}}\relax}%
\newcommand\degr{\arcdeg}%
\newcommand\sun{\odot}%
\usepackage[letterpaper]{geometry}
\usepackage{setspace}
\usepackage[numbers]{natbib}
\usepackage{amsmath}
\usepackage{amssymb}
\usepackage{graphicx}
\usepackage{lipsum}

\makeatletter
\newcommand{\authorfootnotes}{\renewcommand\thefootnote{\@fnsymbol\c@footnote}}%
\makeatother
\begin{document}


\title{A 0.2 $M_{\sun}$ protostar with a Keplerian disk in the very young L1527 IRS system}





\maketitle
\authorfootnotes
\begin{center}
  John J. Tobin\textsuperscript{1}, Lee Hartmann\textsuperscript{2},
  Hsin-Fang Chiang\textsuperscript{3,4}, David J. Wilner\textsuperscript{5} and
  Leslie W. Looney\textsuperscript{3}, 
  Laurent Loinard\textsuperscript{6,7}, Nuria Calvet\textsuperscript{2},
  Paola D'Alessio\textsuperscript{6}  \par \bigskip
\end{center}
\textsuperscript{1}{Hubble Fellow, National Radio Astronomy Observatory, Charlottesville, VA 22903; jtobin@nrao.edu}\\
\textsuperscript{2}{Department of Astronomy, University of Michigan, Ann Arbor, MI 48109}\\
\textsuperscript{3}{Department of Astronomy, University of Illinois, Urbana, IL 61801 }\\
\textsuperscript{4}{Institute for Astronomy and NASA Astrobiology Institute, University
of Hawaii at Manoa, Hilo, HI 96720}\\
\textsuperscript{5}{Harvard-Smithsonian Center for Astrophysics, Cambridge, MA 02138}\\
\textsuperscript{6}{Centro de Radioastronom{\'\i}a y Astrof{\'\i}sica, UNAM, Apartado Postal
3-72 (Xangari), 58089 Morelia, Michoac\'an, M\'exico}\\
\textsuperscript{7}{Max-Planck-Institut f\"ur Radioastronomie, Auf dem H\"ugel 69, 53121 Bonn, Germany}\\

\textbf{In their earliest stages, protostars accrete mass from their surrounding envelopes
through circumstellar disks. Until now, 
the smallest observed protostar/envelope mass ratio was $\sim$2.1\citep{takakuwa2012}. 
The protostar L1527 IRS is thought to be in the 
earliest stages of star formation\citep{andre1993}. Its envelope 
contains $\sim$1 $M_{\sun}$ of material within a $\sim$0.05 pc 
radius\citep{chandler2000,tobin2011}, and earlier observations 
suggested the presence of an edge-on disk\citep{tobin2010b}. Here we report 
observations of dust continuum emission and $^{13}$CO ($J=2\rightarrow1$) 
line emission from the disk around L1527, from which we determine a protostellar mass of
$M_*$ = 0.19 $\pm$ 0.04 $M_{\sun}$ and a protostar/envelope mass ratio 
of $\sim$0.2. We conclude that most of the luminosity is generated 
through the accretion process, with an accretion rate of 
$\sim$6.6$\times$ 10$^{-7}$ $M_{\sun}$ yr$^{-1}$. If it has 
been accreting at that rate through much of its life, 
its age is $\sim$300,000 yr, though theory suggests larger 
accretion rates earlier\citep{foster1993}, so it may be younger. The presence of a 
rotationally--supported disk is confirmed and significantly more
mass may be added to its planet-forming region as well as 
the protostar itself.}

The protostar L1527 IRS (hereafter L1527) in the Taurus cloud, at a distance of about 140 pc,
is one of the nearest Class 0 protostars. This is the earliest phase
 of the star formation process\citep{andre1993} and we show a schematic diagram of a protostellar
system in Figure 1. Observations of dust continuum emission 
toward L1527 were obtained with the Submillimeter Array (SMA)
and Combined Array for Millimeter-wave Astronomy (CARMA)
at $\lambda$ = 870 \mum, 1.3 mm, and 3.4 mm. The 870 \micron\ and 3.4 mm data are shown in Figure 2 with sufficient
resolution to resolve the emission from the disk
midplane, finding it to be extended north-south, like the 3.8 \mum\ dark lane.
The observed disk size is $\sim$180 AU $\pm$ 12 AU in diameter (R$\sim$90 AU), measured from inside the outer contour
plotted in Figure 2; the dust emission appears smaller than the mid-infrared dark lane 
because the lower density outer disk is fainter than the sensitivity limit. 
Other studies did not conclusively detect a disk around L1527 or other Class 0 protostars 
because the spatial resolution was too low to distinguish the disk emission from 
the envelope and/or the disks were too small\citep{jorgensen2007, maury2010}. We estimate a disk 
mass of 0.007$\pm$0.0007 $M_{\sun}$ from the $\lambda$ = 870 \mum\ flux density 
($F_{870\mu m}$=213.6$\pm$8.1 mJy); details
are given in Section 3 of the Supplementary Information. We consider 
this mass a lower limit because the adopted dust opacity is
large\citep{andrews2005}, and we have not accounted for spatial filtering by the interferometer.

We observed the $^{13}$CO ($J=2\rightarrow1$) molecular line transition with CARMA at 1.3 mm.
 This line traces the outflow in most Class 0 protostars\citep{arce2006}; however, 
Figure 3 shows that the $^{13}$CO emission primarily traces the inner envelope and 
disk in L1527. The outflow is detected at velocities less than $\pm$1 \kms, but does not 
affect our analysis, see Section 2 of the Supplementary Information. The $^{13}$CO data have lower resolution
than the 870 \micron\ and 3.4 mm observations (1\arcsec, 140 AU); however, the positional accuracy of line emission is comparable
to the resolution divided by signal-to-noise ratio (typically 5 or higher), enabling us to determine
the location of emission accurately in each velocity channel. Figure 3 shows the
 $^{13}$CO emission from the blue and redshifted components to be on opposite sides of 
protostar, consistent with Keplerian rotation. The emission from the disk is most likely confined
to $\pm$1\arcsec; at larger radii and lower velocities we expect the
flattened envelope to contribute to the kinematics as shown by lower resolution $^{13}$CO ($J=1\rightarrow0$) 
observations\citep{ohashi1997}. The observations shown in Figures 2 and 3 
as a whole provide definitive evidence for a large, rotationally--supported 
disk in this Class 0 protostellar system. Such a disk at this early phase
may be inconsistent with some disk formation models that consider strong magnetic
braking\citep{dapp2010,fromang2008}; however, large disks can form at this stage in models with 
weak magnetic fields\citep{yorke1999,vorobyov2010} or if the magnetic field is not 
aligned with the rotation axis\citep{joos2012}.

Assuming that the disk is rotationally--supported and that
the mass of the protostar is dominant, we can use the position-velocity information
from the molecular line data to determine the protostellar mass.
This has been done for more evolved sources, but not a Class 0 
protostar\citep{simon2000,jorgensen2009}. To determine the mass, we measured the positional
offset of the line emission relative to the protostar (1.3 mm continuum source)
 in each velocity channel (binned to 0.3 \kms) and the position-velocity data are fit with a 
Keplerian rotation curve ($v = (GM/r)^{1/2}$). These data are plotted 
in Figure 4 and least-squares fitting yields a protostellar mass
 of 0.19$\pm$0.04 $M_{\sun}$; the flattening of radius
at velocities less than $\sim$1.5 \kms\ can be attributed to the superposition of rotation 
velocities projected along the line of sight at large radii. We do not expect contributions
from the envelope to affect the fit because its emission is at lower velocities and larger-scales\citep{ohashi1997}.
The edge-on nature of L1527 facilitates this analysis
because the $\sim$85\degr\ inclination\citep{tobin2008,tobin2010b}
does not significantly affect any calculations. Although the model fit in Figure 4 
is simplistic, it highlights the important physics of the problem, and the method is 
consistent with simulated observations of more complicated line radiative transfer models that require many assumptions,
see Section 4 of the Supplementary Information.

Masses have previously been estimated for binary Class 0 protostellar systems using proper motion
measurements at very high resolution\citep{rodriguez2003}, but
with substantial uncertainty due to unconstrained orbital parameters. 
The primary uncertainty in our measurement is whether the
protostellar mass is dominant over the disk/envelope mass at the scales 
we are probing. The disk mass of 0.007 $M_{\sun}$ 
could be up to a factor of a few higher due to opacity uncertainties
and the envelope mass within $R$ = 150 AU is only expected 
to be $\sim$0.01 $M_{\sun}$ since most mass is on large-scales.
If we allow for a factor of four higher disk and envelope masses, they would combine to
contribute at most $\sim$35\% to the total mass.
The kinematic effect of this additional mass should become apparent at
larger disk radii, but the current data are insufficient to distinguish this 
effect. Moreover, the possibility of additional mass
would only cause the protostellar mass to be overestimated.

The ratio of protostellar mass to envelope mass in L1527 is only $\sim$20\%; 
all other protostellar systems with dynamical mass measurements
from disk rotation have protostellar masses greater than twice the surrounding envelope mass\citep{takakuwa2012}.
Therefore, in contrast to these more evolved systems, L1527 will likely 
accumulate significantly more mass. 
Accreting protostars are expected to follow a `birthline' with rising
effective temperature and luminosity with increasing mass; the birthline is also
the starting point of pre-main sequence evolution once the protostar has stopped 
accreting significantly\citep{hartmann1997}.
If L1527 is on the birthline, we can estimate its stellar parameters from the mass. 
We use the birthline model with an accretion
rate of 2$\times$ 10$^{-6}$ $M_{\sun}$ yr$^{-1}$; for a 0.19 $M_{\sun}$ protostar, this model
indicates a radius of 1.7 $R_{\sun}$, effective temperature 
of 3300 K, and a luminosity of 0.3 $L_{\sun}$\citep{hartmann1997}.
This indicates that $\sim$90\% of the 2.75 solar luminosities\citep{tobin2008} 
is supplied by accretion of mass onto the protostar. Thus, the accretion rate of the disk onto the protostar
is $\sim$6.6$\times$ 10$^{-7}$ $M_{\sun}$ yr$^{-1}$, assuming $L_{acc}$ = $GM\dot{M}/R_*$.
If the protostar has been accreting at this rate throughout its life,
its age is only $\sim$300,000 yr, within the expected lifetime of the Class 0 phase\citep{evans2009}.
However, theoretical studies indicate that mass infall/accretion rates may be larger initially and decrease with
time\citep{foster1993}; in addition, protostars are expected to have variable accretion rates\citep{dunham2010}, 
so L1527 could be younger. The dynamical time of the 0.3 pc outflow (red and blue sides)
as measured by a recent survey of Taurus is $\sim$30,000 yr\citep{narayanan2012}. 

The detection of a proto-planetary disk and a measurement of protostellar mass 
have made L1527 one of the best characterized Class 0 protostellar systems known.
Its high accretion rate is nearly a factor of 100
greater than the more evolved pre-main sequence stars with disks; this rate
is high enough to heat the inner disk to temperatures consistent with 
early Solar System conditions\citep{bell2000}. While we cannot say definitively
what L1527 will look like at the end of its formation phase, it does
have the potential to gain as much mass as the sun from its envelope and
it already has a proto-planetary disk with at least 7 Jupiter masses, similar to presumed planet forming
disks\citep{andrews2010}. Therefore, L1527 already has all the elements of a solar system in the making.

Acknowledgments:
We thank the anonymous referees for constructive comments that have
improved the quality of the manuscript.
The authors wish to thank E. Bergin for commenting on the manuscript and
W. Kwon for discussing improvements to the data reduction. 
J. Tobin acknowledges support provided by NASA through Hubble Fellowship 
grant \#HST-HF-51300.01-A awarded by the Space Telescope Science Institute, which is 
operated by the Association of Universities for Research in Astronomy, 
Inc., for NASA, under contract NAS 5-26555. L. H. and J. T. acknowledge partial
support from the University of Michigan. H.-F. C. acknowledges
support from the National Aeronautics and Space
Administration through the NASA Astrobiology Institute under
Cooperative Agreement No. NNA09DA77A issued through the Office of
Space Science. L.W.L. and H.-F. C. acknowledge support from the Laboratory for Astronomical 
Imaging at the University of Illinois and the NSF under grant AST-07-09206.
P. D. acknowledges a grant from PAPIIT-UNAM.
L. L. acknowledges the support of DGAPA, UNAM, CONACyT (M\'exico), 
and the Alexander von Humboldt Stiftung for financial support.
Support for CARMA construction was derived from the states of Illinois, California, and Maryland, 
the James S. McDonnell Foundation, the Gordon and Betty Moore Foundation, the Kenneth T. and 
Eileen L. Norris Foundation, the University of Chicago, the Associates of the California 
Institute of Technology, and the National Science Foundation. Ongoing CARMA development 
and operations are supported by the National Science Foundation under a cooperative 
agreement, and by the CARMA partner universities. The Submillimeter Array is a joint 
project between the Smithsonian Astrophysical Observatory and the Academia Sinica 
Institute of Astronomy and Astrophysics and is funded by the Smithsonian 
Institution and the Academia Sinica. The National Radio Astronomy 
Observatory is a facility of the National Science Foundation 
operated under cooperative agreement by Associated Universities, Inc.

Contributions:
J.T., H.-F.C., D.J.W., and L.W.L. participated in data acquisition 
and reduction. All authors contributed to the data analysis, discussed the results, 
and commented on the manuscript.

Reprints and permissions information is available at www.nature.com/reprints

Competing financial interests:
The authors declare no competing financial interests.

Corresponding Author:
Correspondence and requests for materials should be addressed to John J. Tobin; jtobin@nrao.edu.

\begin{figure}[!ht]
\begin{center}
\includegraphics[scale=1.0]{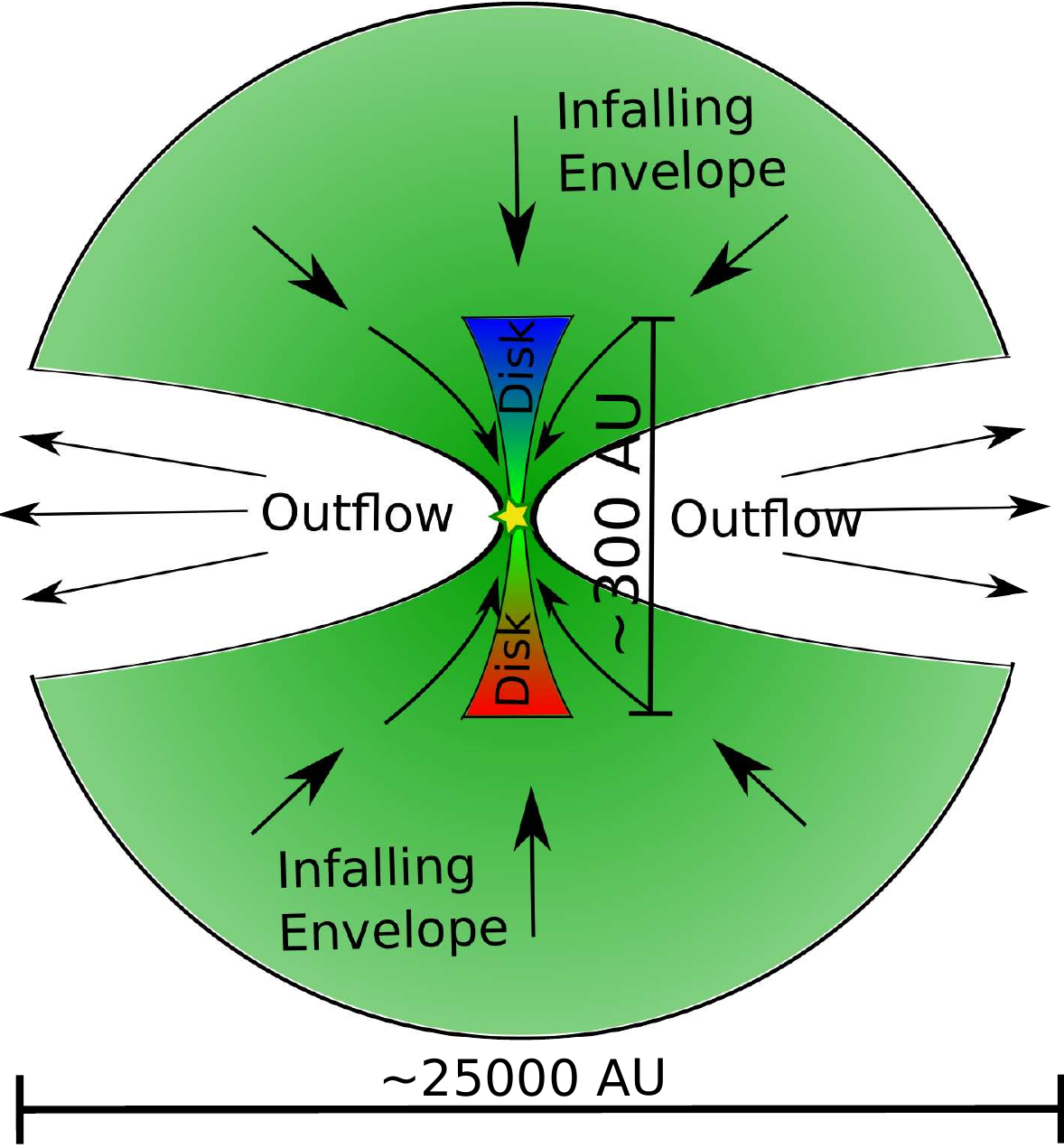}
\end{center}
\caption{Cartoon of a protostellar system rotated to match the orientation of L1527.
 The green highlights the large (R$\sim$12,500 AU) infalling envelope surrounding
the protostar and disk; the envelope geometry on 10,000 AU scales is generally more complex than shown here\citep{tobin2010a}.
Because the infalling material has some net rotation, it falls onto a disk due to conservation of
angular momentum rather than directly onto the protostar.
 The disk is colored with a red to blue velocity gradient to illustrate Keplerian rotation
around the protostar. Mass is transported from the envelope to the disk and then it is accreted through
the disk and onto the protostar. The protostar and disk both work together and drive a bipolar outflow\citep{arce2006} 
which evacuates the polar regions of the envelope. AU is an abbreviation for astronomical unit, the distance from the earth 
to the sun which is 1.496 $\times$ 10$^{13}$ cm.
} 
\label{cartoon}
\end{figure}

\begin{figure}[!ht]
\begin{center}
\includegraphics[scale=0.66,angle=-90]{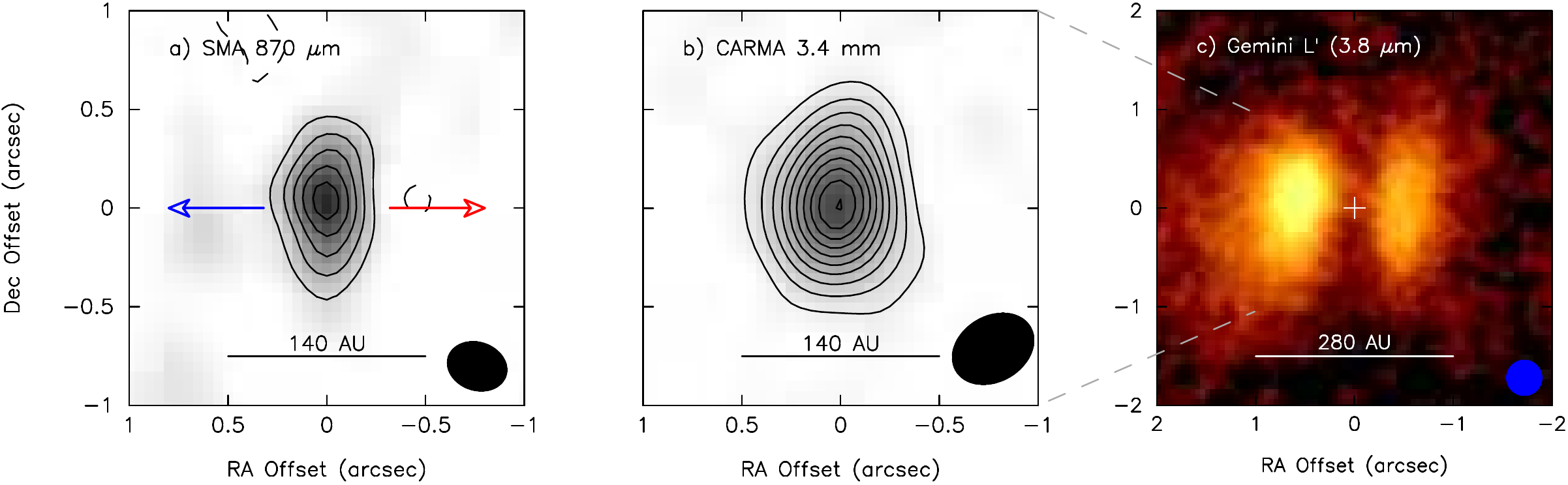}
\end{center}
\caption{Images of the edge-on disk around the protostar L1527.
High-resolution images of L1527 are shown at wavelengths of 870 \micron\ from the SMA
(a), 3.4 mm from CARMA (b), and 3.8 \micron\ from Gemini (c), showing the disk in dust continuum
emission and scattered light. The Gemini image is shown on a larger scale to fully capture
the scattered light features, the dashed gray lines mark the outer edge of the region
shown in the sub/millimeter images.
The sub/millimeter images are elongated in the direction 
of the dark lane shown in panel (c), consistent with
an edge-on disk in this Class 0 protostellar system. 
The outflow direction is indicated by the red and
blue arrows in panel (a), denoting the respective directions of the outflow.
The white cross in panel (c) marks the central position of the disk
from the SMA images. 
The contours in the 870 \micron\ and 3.4 mm images start at 3 times
the noise level and increase at this interval; the noise level 
is 5.0 mJy beam$^{-1}$ and 0.24 mJy beam$^{-1}$ for the SMA and CARMA data respectively.
The ellipses in the lower right corner of each image gives 
the resolution of the observations, approximately 0.25\arcsec, 0.35\arcsec, and
0.35\arcsec\ in the left, middle, and right panels respectively.
}
\end{figure}

\begin{figure}[!ht]
\begin{center}
\includegraphics[scale=0.7,angle=-90]{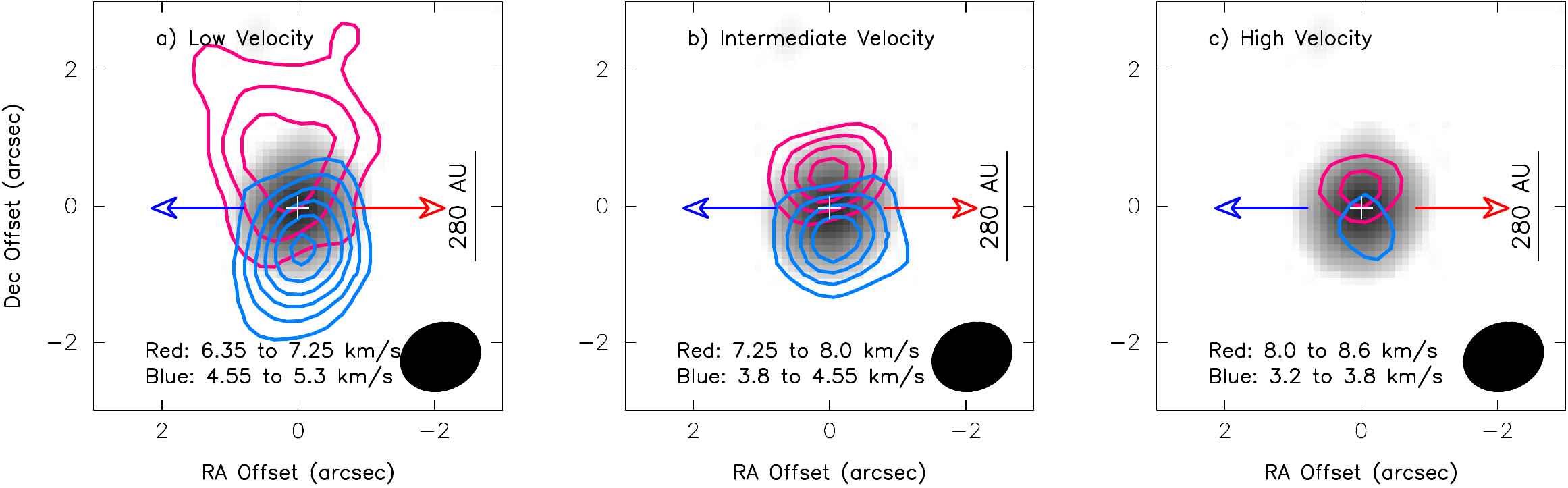}
\end{center}
\caption{$^{13}$CO emission from the disk around L1527 exhibiting a Keplerian rotation signature.
The CARMA 1.3 mm continuum image is shown (grayscale) with the red
and blue contours showing $^{13}$CO emission
integrated at low velocities (a), intermediate velocities (b), and high velocities (c). The white cross in all panels
marks the location of the protostar. The blue and
red-shifted emission centroids show a clear signature 
of rotation on the size-scale of the protostellar disk and no extension of
emission along the outflow. The low-velocity emission likely includes contributions
from the envelope, while the intermediate to high-velocity emission
is dominated by the disk. The low velocity range is from 6.35 \kms\ to 7.25 \kms\ and 4.55\kms\ to 5.3 \kms; the 
intermediate velocity range is from 7.25 \kms\ to 8.0 \kms\ and 3.8 \kms\ to 4.55 \kms; the
high velocity range is from 8.0 \kms\ to 8.6 \kms\ and 3.2 \kms\ to 3.8 \kms.
The contours start and increase in intervals 3 times noise level ($\sigma$)
where $\sigma$ = 0.85 K \kms\ (red) and 0.75 K \kms\ (blue). The angular resolution 
of these data are given by the ellipse in the lower right corners, 1.1\arcsec\ $\times$ 0.95\arcsec.
} 
\label{diskkinematics}
\end{figure}

\begin{figure}[!ht]
\begin{center}
\includegraphics[scale=0.75]{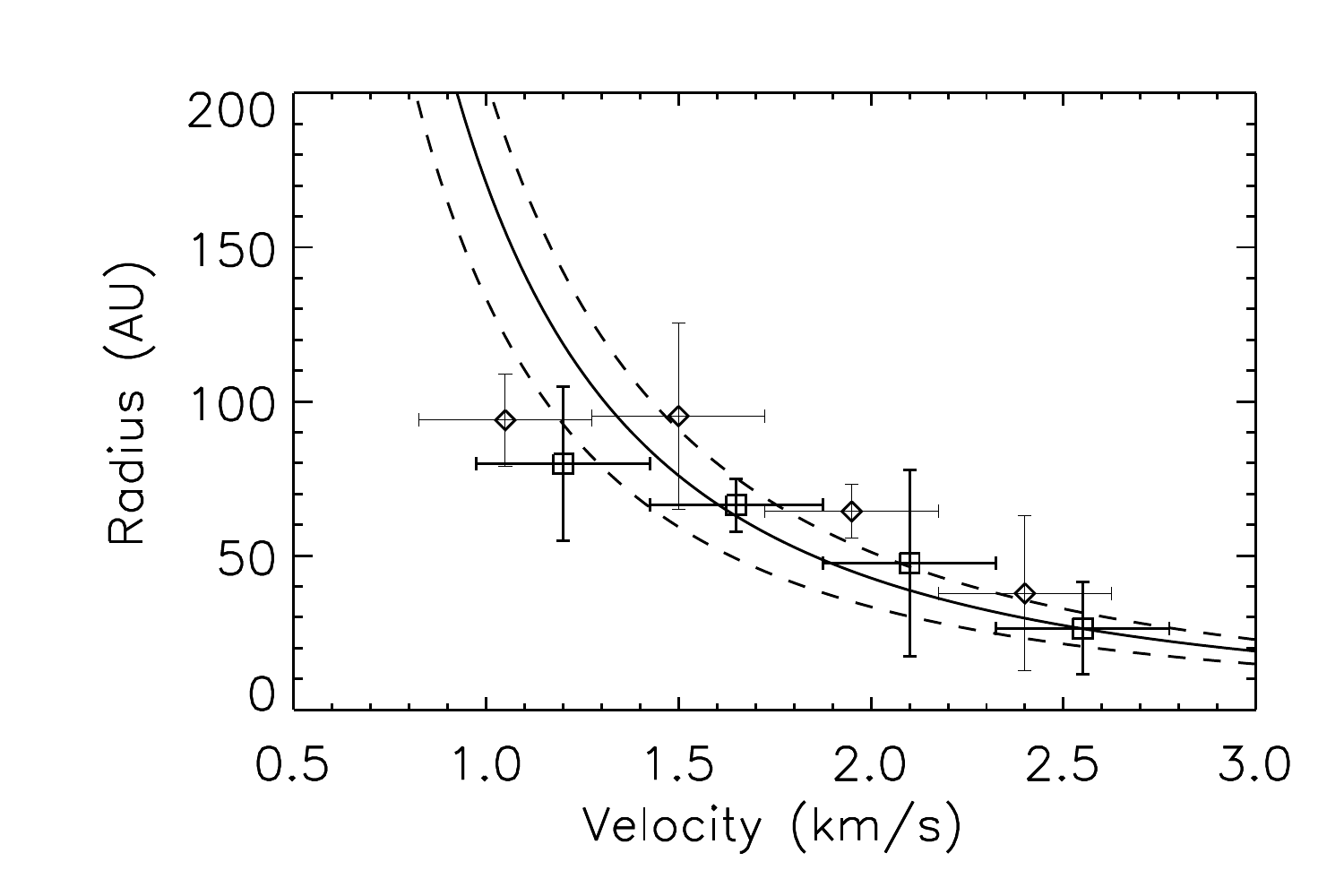}
\end{center}
\caption{Radius versus velocity plots showing the rotation curve derived from the $^{13}$CO emission. Rotationally supported motion 
around a central point mass will have a velocity equal to $(GM/r)^{1/2}$. This function was 
fit to the data, finding a best fitting mass of $M_*$ = 0.19 $\pm$ 0.04 $M_{\sun}$. We used
line radiative transfer models to confirm that this method yields reliable protostellar masses
with simulated data at the same resolution, see Section 4 of the Supplementary Information. The velocity profile
for the best fitting mass is shown as the solid line and the dashed lines show the
1$\sigma$ uncertainty level. The diamonds are data on the south, blue-shifted side of the disk and
the squares are data on the north, red-shifted side of the disk. The uncertainty in the
observed data points is derived from the fixed velocity channel width (x-axis) and error in the radius (y-axis) is
the 1$\sigma$ error derived from adding in quadrature the positional uncertainty of the Gaussian
fit to the $^{13}$CO emission and the error in the position of the protostellar source.
 The flattening of the data points at 
velocities less than 1.5 \kms\ is due to the superposition
of rotation velocities projected to our line of sight at large disk radii and also witnessed
in the data from the line radiative transfer models, see Section 4 of the Supplementary Information.}
\end{figure}

\clearpage
\input{supplementary.tex}

\begin{small}

\bibliographystyle{naturemag}
\bibliography{ms}

\end{small}

\end{document}

%% file: supplementary.tex
\renewcommand{\figurename}{Supplementary Figure}
\setcounter{figure}{0} 
\section{Supplementary Information}

\section{1. Observations}

The data presented in this work were taken with both the Combined Array for Research in 
Millimeter Astronomy (CARMA)\citep{woody2004} and the Submillimeter Array (SMA)\citep{ho2004}, both of
which are sub/millimeter interferometer arrays.
CARMA is a heterogeneous array comprised of six 10.4 m antennas, nine 6 m antennas, and eight 3.5 m antennas at Cedar Flat California
in the Inyo Mountains. The 1.3 mm data were taken in C and D configurations which 
provide angular resolutions of $\sim$1\arcsec\ and $\sim$2.5\arcsec\ respectively and the 
3.4 mm data were taken in A configuration which provides an angular resolution of $\sim$0.35\arcsec.
The SMA is an array of eight 6 m antennas at Mauna Kea and data were taken in the very extended configuration
with a resolution of $\sim$0.25\arcsec.

\subsection{1.1 CARMA 1.3 mm Observations and Data Reduction}
The 1.3 mm C-array observations were conducted on 2012 Jan 29, Feb 04, Feb 06, and Feb 09 using the 10.4 m and 6 m 
antennas. The local oscillator was tuned
to $\nu$=223.01 GHz and four 500 MHz bands were configured for continuum observation. The IF range spans 1 to 9 GHz and 
the receivers operate in dual-sideband (DSB) mode yielding 4 GHz of bandwidth in one polarization. The remaining spectral
windows were configured to observe $^{13}$CO, $^{12}$CO, C$^{18}$O, and H$_2$CO. The Jan 29 data
had excellent phase coherence throughout the track, while the Feb 04, 06, and 09 data had periods of poor phase coherence.
3C111 was observed as the gain calibrator, Uranus was the flux calibrator,
and 3C111 or 3C84 were used as bandpass calibrators. L1527 
was observed in D-array on 2009 Aug 27 with the local oscillator tuned to 
$\nu$=223.1212 GHz, two 500 MHz bands were configured for continuum observation (2GHz DSB)
and one spectral window was centered on the $^{13}$CO line.
3C111 was observed as the gain calibrator, Uranus was the flux calibrator, and 3C84 was the bandpass calibrator.

The data were reduced using the MIRIAD software package \citep{sault1995}. The visibilities were first corrected for
the updated baseline solution and transmission line length correction. Then the amplitudes and phases were examined
for each baseline, flagging the source observations between points where the calibrator had bad amplitudes or extremely 
high phase variance. The data were then bandpass corrected using the \textit{mfcal} task. The absolute flux calibration
was calculated using the \textit{bootflux} task to determine the flux of 3C111 relative to the absolute flux calibrator.
We compared the visibility amplitudes of L1527 in ranges of overlapping uv-coverage to estimate our flux calibration uncertainty.
The data had systematic offsets necessitating rescaling and we estimate an absolute flux 
uncertainty at 1.3 mm of 20\%; however, the rescaling only affects the absolute intensity scale and not the structure
of emission.

The data were then imaged by first computing the inverse Fourier transform of the
data with the \textit{invert} task, weighting by system temperature and creating the dirty map. The dirty map is then CLEANed
using the \textit{mossdi} task down to 1.5$\times$ the RMS noise level 
and the image is restored and convolved with the CLEAN beam by the \textit{restor} task. This procedure was repeated for each channel
in the spectral line image cube and the continuum images were averaged over the entire bandwidth.

\subsection{1.2 CARMA 3.4 mm Observations and Data Reduction}
The 3.4 mm A-array observations of L1527 were taken on 2010 December 02 during 
stable conditions with $\sim$4mm of precipitable water vapor (pwv) using the 10.4 m and 6 m antennas.
The local oscillator was tuned to $\nu$=90.7510 GHz and the correlator was configured for 
4-bit sampling with 8 - 500 MHz moveable spectral windows, measuring continuum emission. 
We arranged windows such that they occupied the 1 to 5 GHz range of IF bandwidth and our
full instantaneous bandwidth was 8 GHz in one polarization. The primary gain calibrator was 3C111,
12.8 degrees away on the sky, 0431+206 was observed as a test source in conjunction with L1527. 3C84 was observed
as the bandpass calibrator, and Neptune was observed as the absolute flux calibrator. We estimate
an absolute flux uncertainty of 10\%. The observations were conducted
in a standard loop, observing 3C111 for 3 minutes, L1527 for 7 minutes, 0431+206 for 1 minute, and then repeating.
Pointing was updated periodically using optical pointing correction and radio pointing was done once during the
track.

The data were reduced in a similar way to the 1.3 mm data; however we also used the
 CARMA paired antenna calibration system (C-PACS)\citep{perez2010} was used for enhanced atmospheric phase correction
in the A configuration. Briefly, the C-PACS calibration works
as follows, the eight 3.5 meter dishes are positioned next to the 10.4 meter and 6 meter antennas
at the longest baselines. While the source(s) is being observed, the 3.5 meter dishes are observing
a nearby calibrator at 30 GHz. The short timescale phase variations during the observation
of the source(s) are corrected from the simultaneous calibrator observations by the
3.5 meter antennas. This correction effectively
reduces the atmospheric decorrelation at the longest baselines, so long as the C-PACS calibrator is 
sufficiently close to the source(s). During the A-array observation of L1527, the C-PACS antennas 
observed the quasar 0428+329 which was 7.4\degr\ away on the sky
with a flux density of 1.9 Jy at 30 GHz.

The data were imaged in the same way as the 1.3 mm data, but we also reconstructed images of the test source 0431+206,
confirming the good 3.4 mm ``seeing.''

\subsection{1.3 Submillimeter Array Observations and Data Reduction}

We observed L1527 with the SMA on the nights of 2011 January 5 and 6 in very extended configuration,
with 7 antennas operating. The first track had excellent phase coherence but the precipitable water vapor (pwv) was $\sim$4.0 mm and
the second track had $\sim$2.0 mm pwv, but worse phase coherence. The local oscillator
was tuned to $\nu$=347.02 GHz and the correlator was used in 4 GHz (DSB) mode.
The correlator was configured for 32 channels per chunk, the coarsest resolution mode
for continuum observations, each chunk is 104 MHz wide and there
are 48 correlator chunks. 3C111 was observed as the gain
calibrator, 0510+180 was observed as a secondary calibrator, 3C279 was the bandpass
calibrator, and Callisto was the absolute flux calibrator.
The data were taken in the following loop: 3C111 was observed for 2.5 minutes, then L1527 was observed for
8 minutes, then 0510+180 was observed for 2.5 minutes, then L1527 again and finishing with 3C111.

The data were reduced using the MIR package, an IDL-based
software package originally developed for the Owens Valley Radio
Observatory and adapted by the SMA group. The data were first corrected for
the updated baseline solutions and then phases and amplitudes on each
baseline were inspected and uncalibrateable data were flagged.
The system temperature correction was then applied to the data and 
the bandpass correction was calculated, trimming three channels
at the edge of each correlator chunk. After bandpass calibration, 
a first-pass gain calibration was performed on 3C111, 0510+180, and Callisto
to measure the calibrator fluxes. The L1527 and 0510+180 data were then 
calibrated for amplitude and phase using only 3C111. We estimate
an absolute flux uncertainty of 10\%. The data were then exported to
MIRIAD and were imaged using the same technique as 
outlined for the CARMA data. 0510+180 was also imaged and 
found to be a point source, confirming the excellent submillimeter seeing 
conditions.

\section{2. Origin of $^{13}$CO Emission}

It is well known that $^{13}$CO frequently traces outflow activity in Class 0 protostars\citep{arce2006}.
However, we have found the $^{13}$CO emission to trace disk rotation in L1527, therefore it is important to rule-out outflow contamination.
Figure S1 shows the $^{13}$CO ($J=2\rightarrow1$) channel maps in 0.3 \kms\ bins at velocities between 3.5 \kms\ and 8.0 \kms.
The wide-angle outflow is apparent as the large ``X'' across the image with substantial spatial filtering; similar in structure
to the outflow cavities traced in \hcop\citep{hoger1998}. The outflow only appears at low velocities ($\pm$ 1.3 \kms\ from line center at 5.9 \kms).
We also detect the flattened envelope structure  previously imaged in $^{13}$CO 
($J=1\rightarrow0$) between velocities of 5.3 and 6.5 \kms \citep{ohashi1997}. 
The emission does not appear as extended as in the previous works due to 
our observations  having higher spatial resolution, thus more spatial filtering of emission.
Moreover, the previous studies had a larger field of view, $\sim$60\arcsec\ versus $\sim$30\arcsec\ in our data.

The signature of disk rotation is found in the compact emission present in all channels with velocities greater than $\pm$0.6 \kms\
from line center. The emission is offset normal to the outflow and the red and blue-shifted emission is on 
opposite sides of the continuum source, orthogonal to the outflow direction. The spatial offsets of the higher velocity 
emission near the protostar are strong evidence that
the $^{13}$CO emission at small-scales is tracing the disk and not outflow emission. There may be some blending of 
outflow/envelope emission with the disk at velocities less than $\pm$0.6 \kms, but outside of this range
the emission from the compact disk is quite distinct.

\section{3. Disk Mass}
If the sub/millimeter emission of L1527 is optically thin and isothermal, then the
mass of the disk can be estimated directly from its measured flux, provided that the temperature is known.
Following previous work on disk masses\citep{andrews2005}, we assume a dust opacity law with a
spectral index ($\beta$) of 1\citep{kwon2009}, normalized to $\kappa_0$=0.035 g cm$^{-2}$ at 850\micron\citep{andrews2005}, assuming a 
dust-to-gas mass ratio of 1:100. We then calculate the mass assuming optically thin emission and constant dust temperature
with the equation
\begin{equation}
M_{dust} = \frac{D^2 F_{\lambda} }{ \kappa_0\left(\frac{ \lambda }{ 850\mu m }\right)^{-\beta}B_{\lambda}(T_{dust}) },
\end{equation}
where $\beta$ = 1, $D$ = 140 pc and $T_{dust}$ is estimated to be 30 K\citep{andrews2005}. 
The integrated flux densities at 870 \micron\ and 3.4 mm were 213.6$\pm$8.1 mJy and 16.9$\pm$1.4 mJy.
This yields a masses of 0.007$\pm$0.0007 $M_{\sun}$ at 870 \micron\ and 0.025 $\pm$ 0.003 $M_{\sun}$ 
with the propagation of statistical error and 10\% absolute calibration uncertainty. 
The uncertainty in this calculation is dominated by our assumptions of the dust mass opacity, $\beta$,
and characteristic dust temperature.  The dust opacity could be could be 
a factor of a few lower\citep{ossenkopf1994,hartmann2008,tobin2010b} which would raise the mass by 
the same factor and a change in dust temperature of $\pm$ 10 K lowers/raises
the disk mass by a factor of $\sim$1.3. The difference in mass between the 870 \micron\
measurement and the 3.4 mm suggests that either the emission is optically thick at 870 \micron\
or the dust opacity has a shallower spectral index than assumed. There is evidence that $\beta$ may
be shallower than 1, if we also consider the recent 
7 mm flux measurement from the EVLA\citep{melis2011}, the spectral slope of emission from 870 \micron, 3.4 mm, and
7 mm is consistent with $F_{\nu}$ $\propto$ $\nu^2$. In the main text, we only list the 870 \micron\ mass
because it is least affected by the assumption of $\beta$ and is directly comparable to previous
work\citep{andrews2005} and exploration of the dust opacity properties of the disk is beyond the scope of this paper.

\section{4. Kinematic Modeling}

In order to better constrain the protostellar mass of L1527, we calculated radiative transfer models
of the $^{13}$CO emission using the LIne Modeling Engine (LIME),
a Monte Carlo spectral line and dust continuum radiative transfer code\citep{brinch2010}.
The temperature and density structure of the envelope and disk is calculated by another radiative transfer
code written by Barbara Whitney that calculates the propagation of luminosity through
the envelope and protostellar disk\citep{whitney2003}.

We have adopted the envelope density and velocity structure
from the rotating collapse model\citep{ulrich1976, cassen1981, tsc1984}.
This density structure is spherically symmetric at large radii, but near centrifugal radius ($R_C$) 
the envelope becomes flattened due to rotation; the envelope density is normalized
to a fiducial density at a radius of 1 AU ($\rho_{1AU}$). The velocities consider infall and
rotation under the assumption that the central protostellar mass is dominant,
in other words we ignore any contributions of mass from the envelope and/or disk. The infalling
envelope becomes rotationally supported at the centrifugal radius ($R_C$), assumed to be equal
to the disk radius. Outflow cavities are also included in this model with
an opening angle of 20$^{\circ}$. The disk model is parametric; we define the radial density profile, 
flaring as a function of disk radius, an initial scale height, and total mass. We also assume a Gaussian vertical disk structure. 
The envelope is assumed to have $\rho_{1AU}$ = 7.25 $\times$ 10$^{-14}$ g cm$^{-3}$, the disk mass is 0.01 $M_{\sun}$, 
the disk radius is 150 AU, the disk flaring $H$ is proportional to $R^{1.3}$ and a scale height at 100 AU of 40 AU.
The infall rate implied by the rotating collapse model is 7 $\times$ 10$^{-6}$
$M_{\sun}$ yr$^{-1}$ when adjusted for the measured protostellar mass. These parameters are selected from disk models
constructed for L1527 that fit the sub/millimeter data, Gemini imaging, and SED\citep{tobin2010b,tobin2012}. Note that the combined
envelope and disk masses for this model are only 0.02 $M_{\sun}$ out to R = 420 AU (3\arcsec).

The $^{13}$CO abundance is assumed to be 1.66 $\times$ 10$^{-5}$ per Hydrogen molecule, where the dust temperature
is greater than 20 K. This assumes a standard CO abundance\cite{lee2004} with a $^{13}$CO/CO ratio of 1/60. We assume
that all CO is frozen onto dust grains at temperatures less than 20 K. The density structure of 
the envelope models does not depend on the protostellar mass, therefore mass can be varied independently of the
envelope. We calculated LIME models for protostellar masses between 0.025 and 0.3 $M_{\sun}$ in 0.0125 $M_{\sun}$
intervals to compare with the observed $^{13}$CO kinematics. On scales less than 1\arcsec\ most of the 
emission is from the disk, while on 2\arcsec\ to 3\arcsec\
scales the emission originates in the inner envelope. However, the rotation velocities of the envelope 
are becoming comparable to the infall velocities on these scales\citep{ulrich1976}.  

Each model was run through a simulated
observation with the \textit{simdata} task of the CASA software package\footnote[1]{http://casa.nrao.edu}. This 
calculates the Fourier transform of the emission in each velocity channel, samples it with uv-coverage similar
to what is provided by the CARMA C and D configurations at 1.3 mm, and then reconstructs the image from the inverse
Fourier transform and the CLEAN algorithm. 

We compared the observations to the models in two ways. The first method was the same as used in the main paper to 
fit the mass of the protostar. This was done by calculating the centroid of the line emission offset from the protostar at a given velocity
within a velocity range of 0.5 to 3.0 \kms. This enabled us to verify that the method used to derive the mass of L1527 was reliable, using models
with a known protostellar mass.  The models masses derived from the least-squares fits versus actual model
masses are given in Table S1, showing that a model with an actual mass of 0.225 $M_{\sun}$ has a 
mass fit of $M_{fit}$ = 0.19 $M_{\sun}$, slightly underestimating 
the true mass of the model. Thus, the mass of L1527 could be closer to 0.225 $M_{\sun}$;
however, this is already within the statistical uncertainty of the least-squares fit. 
The model data points are overlaid on the observations in Figure S2 and
agree very well with the positions and velocities of the observed data. 
Position-velocity diagrams of the $^{13}$CO emission for L1527 and the $M_*$ = 0.225 $M_{\sun}$
model are shown in Figure S3. While the data are noisy and have some emission asymmetries, the plots qualitatively agree.
This indicates that a disk and a infalling/rotating envelope\citep{ulrich1976,cassen1981}
describe the velocity structure quite well at these small-scales.

The second method we used to derive the mass of L1527 involved subtracting the model 
emission from the data in each velocity channel and calculating 
\begin{equation}
\chi^2 = \sum_{\alpha,\delta,v}^{N}\frac{(I_{obs}(\alpha,\delta,v) - I_{model}(\alpha,\delta,v)^2}{\sigma_i^2}.
\end{equation}
 $\chi^2$ is calculated within a 2\arcsec\ ($\alpha$) $\times$ 4\arcsec\ ($\delta$)
region centered on the protostar in the velocity ranges between 3.5 \kms\ $<$ V $<$ 4.7 \kms\ and 7.1 $<$ V $<$ 8.6 \kms.
This subregion of the data cube contains most of the emission from the protostar and disk as well as 
selecting out the velocity range near the systemic velocity that is affected by opacity and spatial filtering.
The noise can be described by a 
Gaussian, but it is correlated on the scale of 1\arcsec. Because the image is sampled with 0.2\arcsec\ pixels,
the degrees of freedom are reduced by a factor of 25, yielding 1056 degrees of freedom from the image region and velocity
channels.

Unfortunately, the mass is not tightly constrained by the channel map analysis, likely due to abundance variations
of $^{13}$CO and the low spatial resolution of the data compared to the size scale of emission. The reduced $\chi^2$
values for each mass are given in Table S2, showing that masses between 0.075 and 0.175 $M_{\sun}$ are 
equally probable (or improbable), with 0.125 $M_{\sun}$ having the lowest reduced $\chi^2$. This comparison
of the models to the data indicate a low mass, similar to the first method, giving further confidence
that our first method using least-squares fitting is reliable.



\begin{figure}[!ht]
\begin{center}
\includegraphics[angle=-90, scale=0.8]{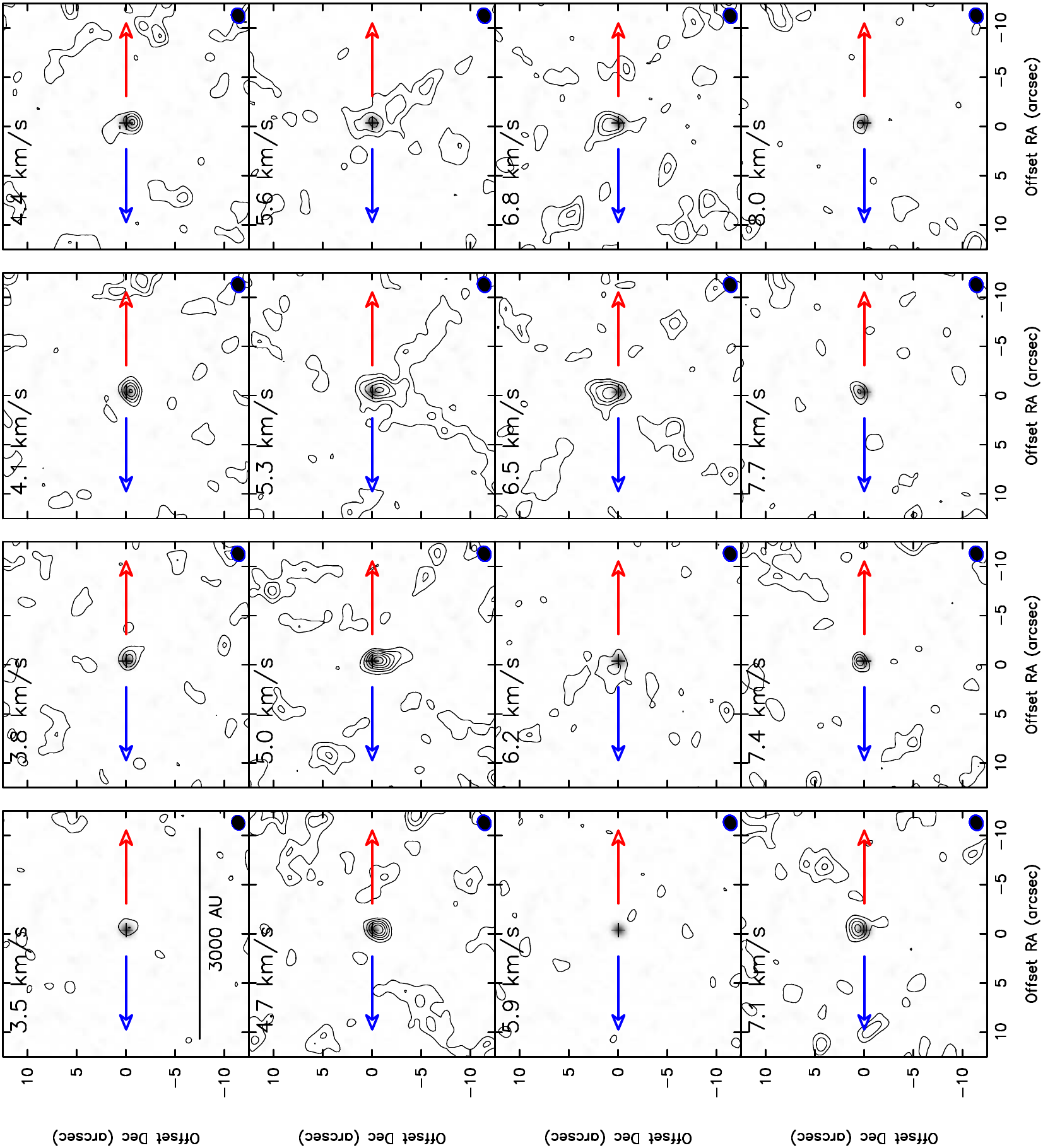}
\end{center}
\caption{Channel maps of $^{13}$CO emission in L1527. The velocity channels are 0.3 \kms\ wide
and span the entire velocity range of emission detected in L1527. The line center velocity is 5.9 \kms\
where emission is absent due to spatial filtering and optical depth. At velocities within $\pm$1.3 \kms,
the outflow is evident as the large ``X'' feature and the protostellar envelope is also visible in this velocity
range. At higher velocities the emission is more compact and the red and blue-shifted emission appears on opposite sides, in the
same direction as the disk and orthogonal to the outflow. This is strong evidence for
the $^{13}$CO to be tracing disk rotation. The contours start at 1.8 K (2$\sigma$) and increase in units
of 2$\sigma$. The angular resolution of the channel maps has been tapered to 1.6\arcsec\ $\times$ 1.3\arcsec\ to
better emphasize the large-scale outflow features.
}
\end{figure}

\begin{figure}[!ht]
\begin{center}
\includegraphics[scale=0.75]{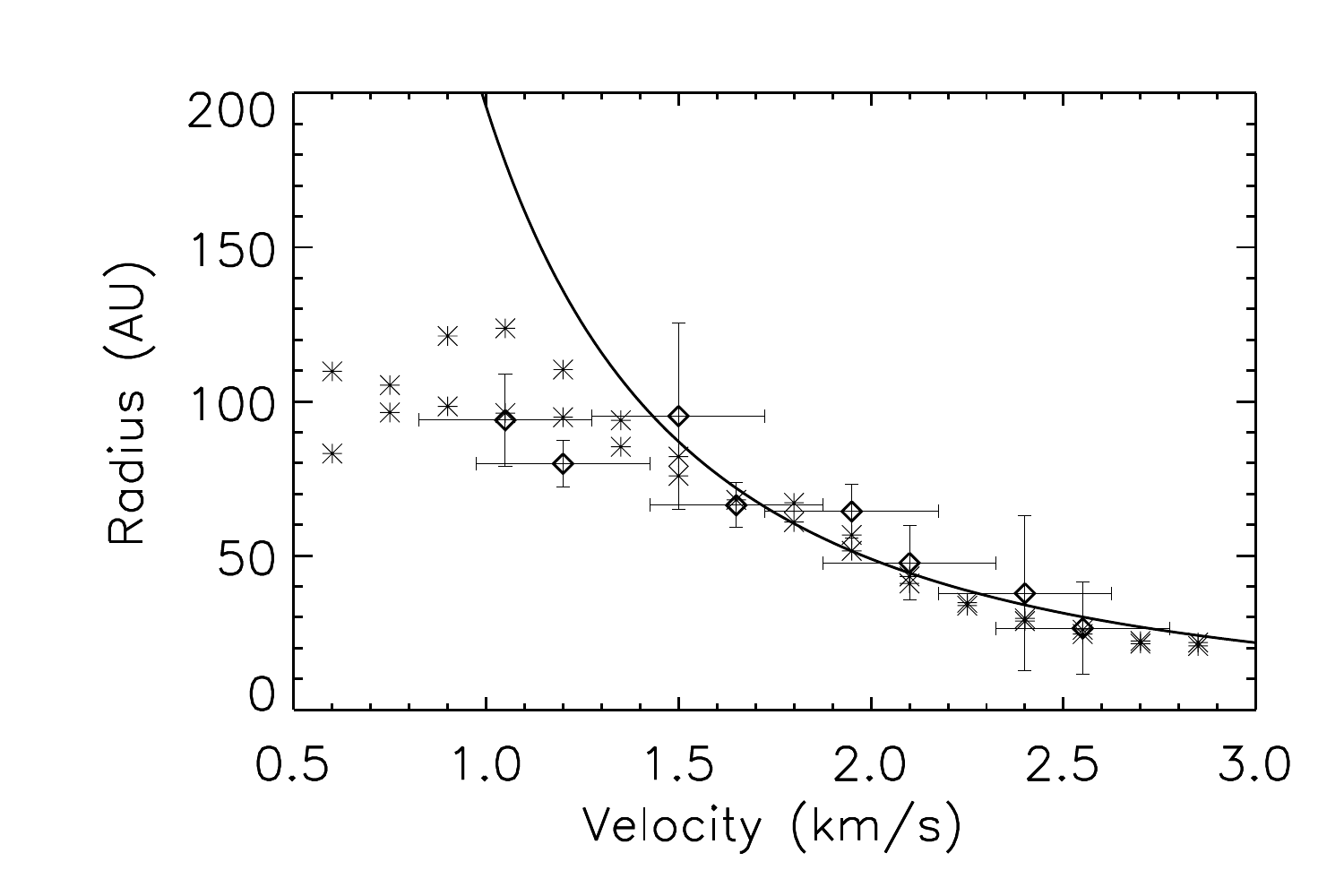}

\end{center}
\caption{Radius versus Velocity plots of $^{13}$CO emission for the L1527 observations (diamonds) the
same measurements taken on the molecular line model (asterisks). Rotationally supported motion 
around a central point mass will have a velocity equal to $(GM/r)^{1/2}$. This function was 
fit to the observational and model data, finding a best fitting mass of $M_*$ = 0.19 $\pm$ 0.04 $M_{\sun}$. The velocity profile
of the best fitting mass to the model and data is shown as the solid line.
The actual mass of the model is 0.225 $M_{\sun}$.  The uncertainty in the
observed data is derived from the channel width and the positional uncertainty of the Gaussian
fit to the emission and the error in the position of the protostellar source.
 The data points of the model are entirely consistent with the observed data, including
the flattening of radius at velocities less than 1.5 \kms. This flattening is due to the superposition
of rotation velocities projected to our line of sight at large radii.}
\end{figure}

\begin{figure}[!ht]
\begin{center}
\includegraphics[scale=0.7,angle=-90]{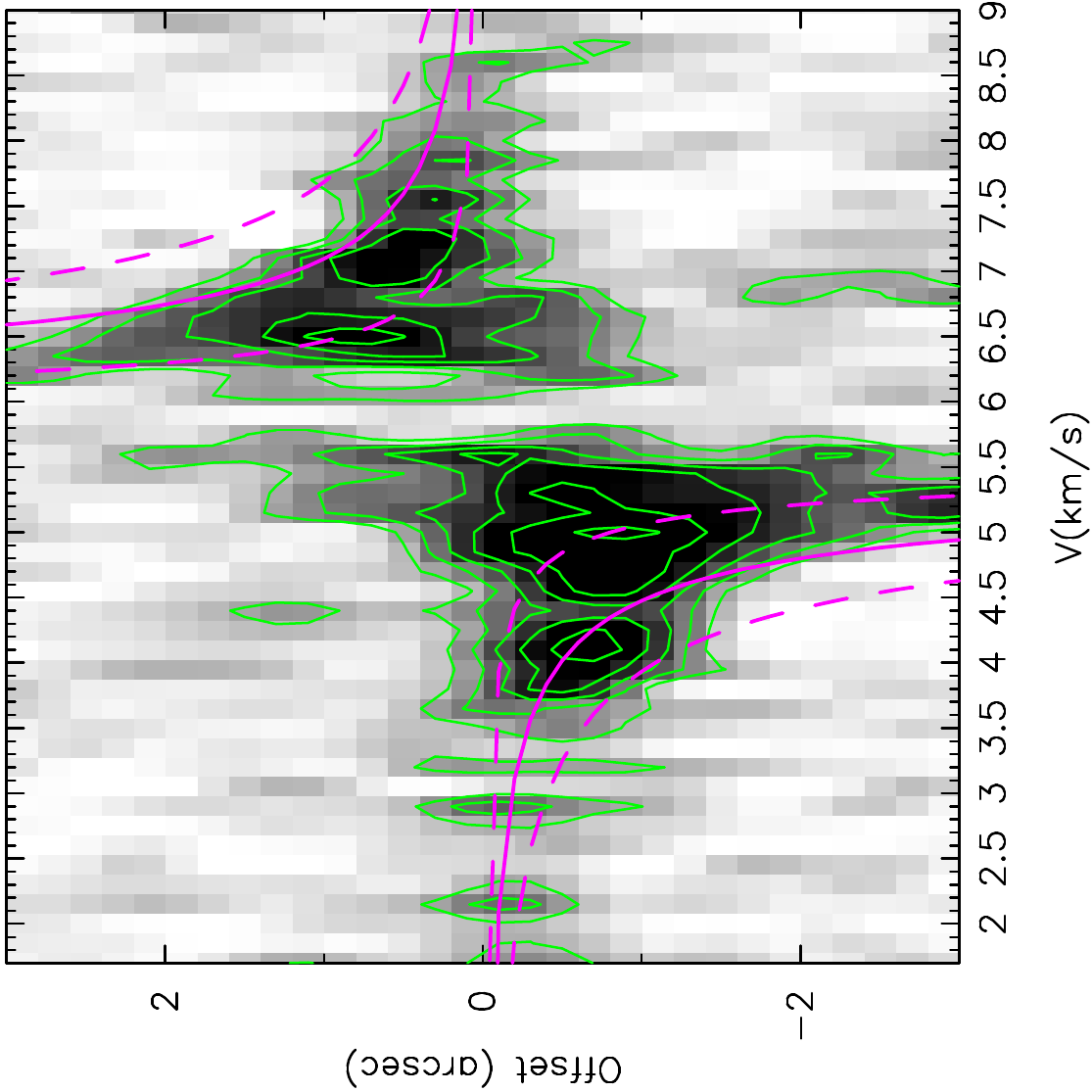}
\includegraphics[scale=0.7,angle=-90]{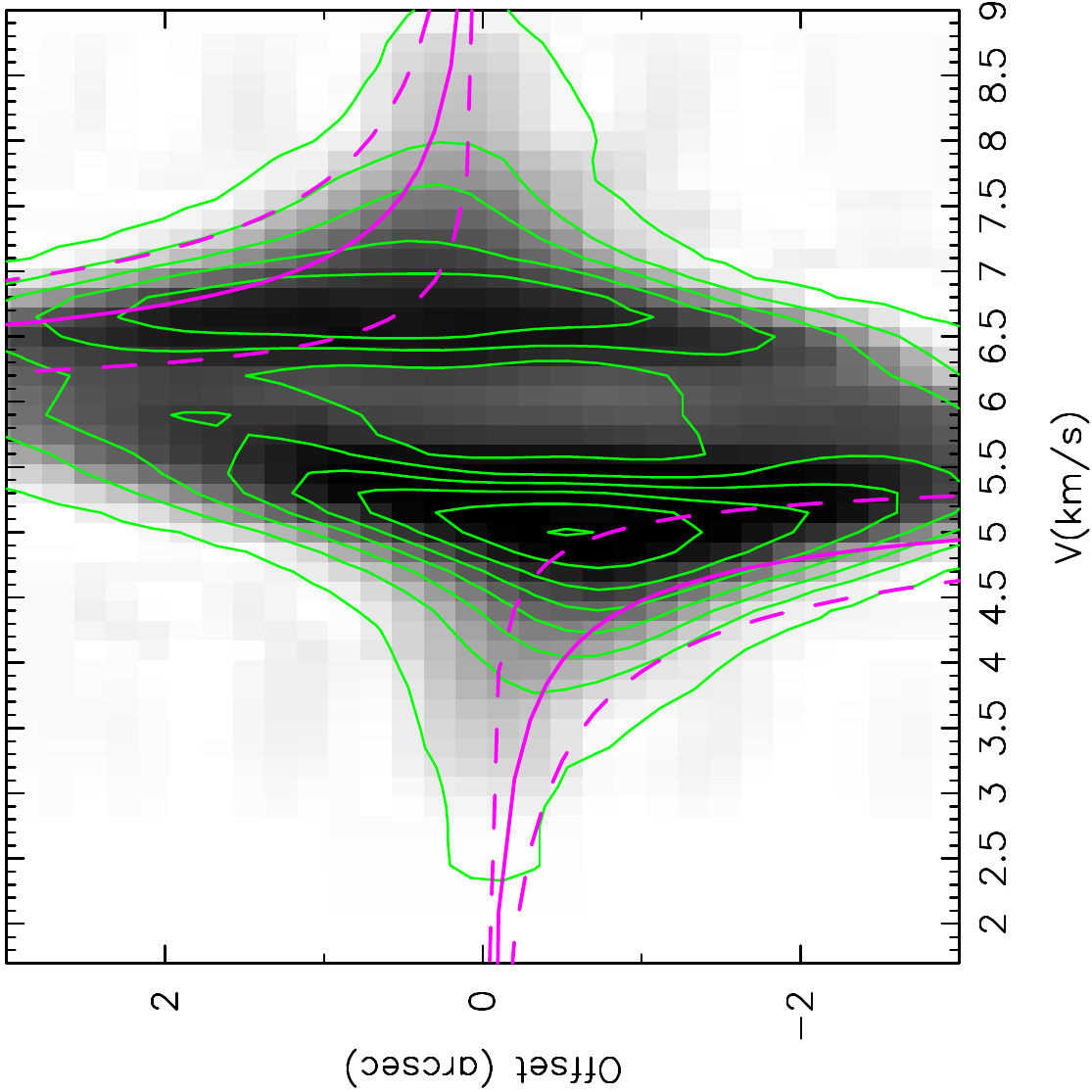}
\end{center}
\caption{Position-velocity diagrams of $^{13}$CO in L1527 (left) and the  model which agrees with 
our best-fitting mass (M = 0.225 $M_{\sun}$) (right) taken along the north-south axis of the disk/envelope.
The PV structure of the observation closely resembles what is expected for rotating collapse, the
deficit emission at 5.9 \kms\ is due to a combination of spatial filtering and self-absorption.
The rotation of the disk is evident at the higher velocities within $\pm$1\arcsec. 
The solid magenta lines are the
Keplerian velocities for a 0.225 $M_{\sun}$ object and the dashed magenta lines are Keplerian rotation
curves for masses of 0.05 and 0.5 $M_{\sun}$. The emission at radii greater than 1\arcsec\
is at a lower velocity than the best fitting mass because this is outside the likely disk radius and
not rotationally supported; this emission dominated by infall velocities.
The contours for the $^{13}$CO data start at 2.5 K (5$\sigma$) and the contours on the left
start at 0.5 K.}
\end{figure}

\clearpage
\begin{center}
Supplementary Table 1. Least-Squares Mass Fits on the Models \\
\vspace{0.1cm}
\begin{tabular}{cc}
\hline
\hline
Actual Model Mass & Fit Model Mass\\ 
 ($M_{\sun}$)  & ($M_{\sun}$)\\
\hline
0.025 & 0.06\\
0.05 & 0.04\\
0.075 & 0.11\\
0.1 & 0.11\\
0.125 & 0.13\\
0.15 & 0.14\\
0.175 & 0.15\\
0.2 & 0.17\\
0.225 & 0.19\\
0.25 & 0.21\\
0.275 & 0.22\\
0.3 & 0.24\\
\hline
\hline
\end{tabular}

The statistical uncertainties in the mass fit are typically 0.03 $M_{\sun}$ and the
data lack enough resolution to constrain masses less than 0.05 $M_{\sun}$.\\
\end{center}

\begin{center}
Supplementary Table 2. Results of Channel Map Fitting to the Data \\
\vspace{0.1cm}
\begin{tabular}{cc}
\hline
\hline
Model Mass & Reduced $\chi^2$\\ 
 ($M_{\sun}$)  & \\
\hline
0.025 & 24.7\\
0.05 & 20.9\\
0.075 & 17.2\\
0.1 & 14.5\\
0.125 & 13.8\\
0.15 & 14.3\\
0.175 & 15.9\\
0.2 & 18.5\\
0.225 & 22.8\\
0.25 & 27.2\\
0.275 & 31.4\\
0.3 & 37.4\\
\hline
\hline
\end{tabular}
\end{center}